# Corruption Determinants, Geography, and Model Uncertainty

Sajad Rahimian
MA Student in Economics, University of Saskatchewan
sar291@usask.ca

## 1. Abstract

This paper aims to identify the robust determinants of corruption after integrating out the effects of spatial spillovers in corruption levels between countries. In other words, we want to specify which variables play the most critical role in determining the corruption levels after accounting for the effects that neighbouring countries have on each other. We collected the annual data of 115 countries over the 1985-2015 period and used the averaged values to conduct our empirical analysis. Among 39 predictors of corruption, our spatial BMA models identify Rule of Law as the most persistent determinant of corruption.

## 2. Introduction

Corruption is an area of contention, both in the public sphere and scientific community. Understanding its manifestations, causes, consequences, and crafting efficient policies to curb it has elicited numerous efforts from social scientists, international bodies, and governments. More specifically, the ongoing debate over its underlying driving factors has resulted in a variety of theories and contrasting empirical findings.

On top of the measurement complications and econometrics hurdles, abound theories in literature have led to uncertainty regarding the most relevant covariates of corruption (Jetter and Parmeter, 2018). This ambiguity often referred to as "model uncertainty", has prompted a handful of researchers to examine wide arrays of candidate explanatory variables to uncover the robust determinants of corruption.

In one of the earliest and well-received studies of this kind, Treisman (2000) provides a comprehensive quantitative assessment of potential determinants of corruption. By employing OLS regression analysis, his research unveils the positive role of Protestant traditions, British rule, economic development, imports, and enduring democracy in keeping corruption under control. Moreover, he cautions about the adverse effects of federalism in exacerbating corruption. In a similar vein, Serra (2006) applies Extreme-Bounds Analysis to a set of 28 corruption determinants. Her research identifies economic development, long-lived democracy, Protestant population, political instability, and colonial heritage as significant antecedents of corruption. In a recent attempt, Jetter and Parmeter (2018) have tried to pinpoint the persistent regressors of corruption in a Bayesian Model Averaging (BMA) framework. Rule of law, government effectiveness, urbanization, the share of women in parliament, and primary schooling emerge as the robust predictors of corruption among 36 independent variables. Our study aims to contribute to this field of knowledge and builds on the work of Treisman (2000), Serra (2006), and Jetter and Parmeter (2018). We explicitly address a shortcoming of these three studies, that is, their negligence over an important attribute of corruption, namely: spatial spillovers.

A growing body of research highlights the contagious nature of corruption within and between countries (Borsky and Kalkschmied, 2019; Correa et al., 2016; Lopez-Valcarcel et al., 2017; You and Nie, 2017). Corruption could diffuse across adjoining regions through a variety of transmission channels. First, economic exchange and particularly trade ties by altering the balance of political power or changing the domestic agents' beliefs, expectations, and preferences, could result in institutional change for better or worse. Second, cultural or linguistic affinities and migration provide a level ground for exchanging ideas, beliefs, and expectations across proximate geographic units. Notably, migrants take on economic, social, and political roles in host societies and contribute to upholding or deteriorating the predominant institutions. Lastly, political relationships could exert an influence on institutional quality. Governments at national and subnational levels might choose to adopt institutions from neighboring governments to pave the way for more political engagement or economic cooperation (Borsky and Kalkschmied, 2019).

Failure to incorporating such spatial dependencies in empirical analyses of corruption could result in omitted variables bias, spurious coefficient estimates, and probably erroneous conclusions. But entering spatial correlations into econometric models invokes certain complexities, among which, the choice of spatial weight matrices is of critical importance. Misspecification of contiguity matrices could have dire repercussions for the credibility of research results. Unfortunately, despite their importance, literature usually provides little guidance on the true pattern of geographical relationships and the structure of spatial weight matrices. This implies an uncertainty regarding the choice of the best contiguity matrix in empirical spatial studies and renders a challenge to empirical researchers.

To single out the robust regressors of corruption from a comprehensive list of potential determinants, we use the methodology developed by Crespo Cuaresma and Feldkircher (2013). Their approach is well-suited to serve our objective. Their method filters out spatial correlations in a BMA context and not only addresses model uncertainty but also helps to overcome the uncertainty regarding the best spatial weight matrix. Thus, the novelty of our study is twofold. First, by utilizing the spatial BMA technique proposed by Crespo Cuaresma and Feldkircher (2013), we provide a new perspective on the robust determinants of corruption after integrating out the effects of spatial spillovers. Second, we overcome the uncertainty stemming from the choice of spatial linkages in a BMA framework. We enter several contiguity matrices into our analysis and choose the one that best describes the neighborhood relationships.

## 3. Literature review

Following Jetter and Parmeter's (2018) classification, we arrange the likely regressors of corruption into four categories (i.e. institutional, economic, cultural, and geographical groups) and briefly survey the literature accordingly. It should be noted that there is no unanimous agreement among scholars regarding the effects of our explanatory variables on corruption. Covering the divergent arguments and conclusions in literature requires an exhaustive literature review and is beyond the scope of this paper. Here, we suffice to outline the prevalent viewpoint about the impact of each of the explanatory variables on corruption.

### 3.1. Institutional determinants

Institutional factors are the first category of interest. We start with government size and effectiveness. Generally, an increase in government size is perceived to aggravate corruption (Alesina and Angeletos, 2005). Asongu (2013) and Dreher et al. (2009) conclude that government effectiveness deters corruption. Further, Nur-Tegin and Czap (2012) reveal that democracy and corruption are negatively associated. Additionally, Rock (2009) provides empirical evidence that the durability of democracy matters and corruption decreases after 10-12 years since the inception of democratic governments. Moreover, the positive effects of political rights (Shen and Williamson, 2005), property rights (Lima and Delen, 2019), rule of law (Herzfeld and Weiss, 2003), decentralization (Lessmann and Markwardt, 2010), and press freedom (Brunetti and Weder, 2003) in discouraging corrupt activities are well documented in the literature. Finally, Serra's (2006) findings underscore the importance of "legal cultures."

### 3.2. Economic determinants

Besides institutional factors, many researchers have scrutinized the association between economic variables and corruption. Higher GDP per capita is found to be a deterrent to corruption by many studies (Treisman, 2000; Serra, 2006). Trade, imports, and economic globalization are at the center of various analyses and are widely believed to alleviate corruption (Ades and Di Tella, 1999; Badinger and Nindl, 2014). Studies on the FDI-corruption nexus suggest confounding results and are yet to reach a definitive verdict (Larraín and Tavares, 2004; Pinto and Zhu, 2016). Further, a bulk of the literature focuses on the natural resources and the so-called "resource curse" phenomenon. For instance, Kolstad and Søreide (2009) document the harmful consequences of resources abundance in terms of rampant corruption. Moreover, empirical findings lend credit to the view that greater urbanization hinders corruption (Billger and Goel, 2009; Goel and Nelson, 2010). Lastly, Botero et al. (2013) and Uslaner and Rothstein (2016) elaborate on the favorable effects of better education on constraining corruption.

### 3.3. Cultural determinants

In addition to institutional and economic predictors, cultural elements influence corruption levels. We start our discussion with the population. Knack and Azfar (2003) expand on the complex relationship between population and corruption and conclude that no clear relationship could be detected between the two phenomena. Furthermore, a ubiquitous empirical finding indicates that a higher fraction of Protestants in a country relates to lower levels of corruption (Gokcekus, 2008). Ethnolinguistic fractionalization is the next key cultural determinant and previous research finds it to aggravate corruption (Shen and Williamson, 2005). Moreover, women's participation in politics could help to bolster up anti-corruption efforts and reduce corruption (Jha and Sarangi, 2018). Finally, countries' colonial heritages might have implications for their contemporary corruption. This proposition is buttressed by the empirical findings that affirm former British colonies exhibit lower corruption (Serra, 2006; Treisman, 2000).

### 3.4. Geographical determinants

Geographical attributes constitute the last group of corruption determinants. We add latitude to our variables list based on the prior work by Dreher et al. (2009) and La Porta et al. (1999). We take our cue from Oberdabernig et al. (2018) and include regional dummy variables in our analysis.

## 4. Methodology

### 4.1. General framework

We follow the methodology put forward by Crespo Cuaresma and Feldkircher (2013) to conduct our spatial BMA analysis. This section provides a summary of Crespo Cuaresma and Feldkircher's (2013) underlying idea of spatial BMA. The *R* package *spatBMS* was used to carry out the models.

In a cross-sectional regression for $N$ regions, the dependent variable $y$ is a column vector with $N$ dimensions. Suppose that the pattern of spatial spillovers falls within the class of spatial autoregressive (SAR) models:

$$y = \rho \boldsymbol{W} y + \boldsymbol{X}_k \overrightarrow{\lambda_k} + \sigma\varepsilon \tag{1}$$

Where $\boldsymbol{X}_k$is an $N \times k$ matrix and $k$ represents the number of explanatory variables. $\overrightarrow{\lambda_k}$ is the vector of parameters with $k$ elements corresponding to the covariates in $\boldsymbol{X}_k$. $\boldsymbol{W}$ is the spatial weight matrix and defines the pattern of spatial dependencies. The spillover parameter $\rho$ reflects the degree of spatial autocorrelation.

Two types of uncertainty emanate from equation (1). In addition to the uncertainty about the independent variables to be included in $\boldsymbol{X}_k$, the true structure of $\boldsymbol{W}$ is unknown to researchers in most applications. Addressing the latter uncertainty is of particular importance since statistical inference from a SAR model is conditional on the exogenously determined spatial weight matrix $\boldsymbol{W}$.

**4.2. Spatial filtering**

Crespo Cuaresma and Feldkircher (2013) devise a procedure to overcome these two uncertainties in a Bayesian framework. They implement the eigenvector spatial filtering method suggested by Griffith (2000) and Tiefelsdorf and Griffith (2007) to free the regression residuals from spatial autocorrelation patterns. They add the eigenvectors $\{e_i\}$, extracted from the decomposition of a transformed $\boldsymbol{W}$ matrix, to the equation (1) as independent variables:

$$y = \sum_{i=1}^{E} \gamma_i \overrightarrow{e_i} + \boldsymbol{X}_k \overrightarrow{\lambda_k} + \sigma\varepsilon \tag{2}$$

Where, $E$ is the set of required eigenvectors ($\boldsymbol{E}$ = $\{\overrightarrow{e_1}, \ldots, \overrightarrow{e_E}\}$) identified by an algorithm proposed by Tiefelsdorf and Griffith (2007) and "each eigenvector $\overrightarrow{e_i}$ spans one of the spatial dimensions" (Crespo Cuaresma and Feldkircher, 2013). Incorporating eigenvectors into the regression takes care of the spatial dependencies in the residuals. This enables researchers to estimate equation (2) instead of equation (1).

**4.3. Uncertain spatial spillovers in the BMA framework**

By exploiting spatial filtering techniques, BMA could give estimates of the parameters when there are uncertainties about the nature of $\boldsymbol{W}$ and $\boldsymbol{X}_k$. Denote the number of spatial weight matrices $\boldsymbol{W}_z$ with $\boldsymbol{Z}$, (z = 1, …, Z) and each matrix has its associated set of eigenvectors $\boldsymbol{E}_z$. Therefore, the model space ($\mathcal{M}$) consists of $2^k \times Z$ possible specifications, $\mathcal{M}$ = $\{M_1^1, \ldots, M_{2^k}^1, M_1^2, \ldots, M_{2^k}^2, M_1^Z, \ldots, M_{2^k}^Z\}$. The parameter vector ($\theta_k^z$) of a particular model ($M_k^z$) consists of $\theta_k^z = (\alpha, \lambda_k, \gamma_z)$ where $\alpha$ is the intercept term included in all models, $\lambda_k$ is the coefficients on the covariates included in the model and $\gamma_z$ is the coefficients on the set of eigenvectors $\boldsymbol{E}_z$ representing $\boldsymbol{W}_z$. The posterior distribution of a specific coefficient ($\beta$) could be expressed as:

$$p(\beta \mid y) = \sum_{j=1}^{2^k} \sum_{z=1}^{Z} p(\beta \mid M_j^z, y) p(M_j^z \mid y) \tag{3}$$

And

$$p(M_j^z \mid y) = \frac{p(y \mid M_j^z)\, \bar{p}(M_j^z)}{\sum_{j=1}^{2^k} \sum_{z=1}^{Z} p(y \mid M_j^z)\, \bar{p}(M_j^z)} \tag{4}$$

$y$ denotes the data, $p(M_j^z \mid y)$and $\bar{p}(M_j^z)$ are the posterior model probability and the prior distribution of model $M_j^z$ respectively, and $p(y \mid M_j^z)$ is the integrated likelihood. Because of the enormous size of the model space ($\mathcal{M}$) and the high correlation among eigenvectors in many cases, Crespo Cuaresma and Feldkircher (2013) design a special Markov chain Monte Carlo algorithm to explore the model space in pursuit of best models. Each step of their sampler comprises two sub-steps. First, the sampler chooses between two models that have similar $\boldsymbol{W}_z$ matrices (and the same set of eigenvectors $\boldsymbol{E}_z$) but differ in only one explanatory variable based on a specific acceptance probability. In the second sub-step, two models with similar sets of independent variables and two different sets of eigenvectors (derived from two different $\boldsymbol{W}$ matrices) are compared based on a specific acceptance probability. Each iteration of the Markov chain comprises of these two sub-steps and the BMA statistics are calculated accordingly. Specifically, The PIP of a particular covariate $l$ is defined as:

$$PIP_l = \sum_{j=1}^{2^k} \sum_{z=1}^{Z} p(M_j^z \text{ such that } \lambda_l \neq 0 \mid y) \quad (5)$$

## 5. Estimation

### 5.1. Data

Our data is structured as cross-section and consists of 115 countries. I collected the data of 41 variables (two dependent variables and 39 independents). Although our variables list is not similar to their research completely, I chose the variables primarily based on the Jetter and Parmeter's (2018) study. Table 9 gives information on data sources and definitions. I used the data of *CPI* and *Control of Corruption* indices in 2015 as our dependent variables. To alleviate endogeneity concerns, I followed Oberdabernig et al. (2018) and averaged the annual observations of all independent variables from 1985 to 2015. Moreover, similar to Oberdabernig et al. (2018) I standardized all the variables (both dependent and independent variables, except for dummy variables) so that our variables have a mean of zero and standard deviation of 1. This makes the coefficients of the explanatory variables directly comparable.

### 5.2. General econometric framework

Our econometric approach is based on Crespo Cuaresma and Feldkircher's (2013) paper. Following these authors, we assume that corruption spillovers have a spatial autoregressive (SAR) pattern. This means that the level of corruption in the country *i* is affected by the corruption levels in its neighboring countries $j_1$, $j_2$, $j_3$, …, $j_n$. To attain our objective, I employed the Spatial Bayesian Model Averaging approach introduced by Crespo Cuaresma and Feldkircher (2013). Their codes and estimation routines are available by the R package *spatBMS*. Please read the *spatBMS* tutorial for more information. All the figures and tables presented in this report are produced using the R software.

### 5.3. Spatial weight matrices

To account for different possible patterns of spatial spillovers between corruption levels across sample countries, I used the following contiguity matrices in the spatial BMA models: *queen*, *1500 kilometer*, *inverse 1500 kilometer*, *4 nearest neighbors*, *6 nearest neighbors*, and *8 nearest neighbors*. I used GeoDa to create the contiguity matrices and then exported them to R. I will introduce spatial weight matrices in the following lines and elaborate on the modifications I made to three of these matrices (i.e. *queen*, *1500 kilometer*, and *inverse 1500 kilometer*). It should be mentioned that similar to Oberdabernig et al. (2018), I calculated the distances between the sample countries based on the great-circle distances between their capital cities (I took the curvature of Earth into account). Moreover, when executing the spatial BMA models, I row-standardized the spatial weight matrices which is a common practice in spatial analyses.

*Queen*: In principle, it is a 115*115 matrix with binary elements. For each element, if the corresponding countries (row and column) have a common vertex the element equals one, otherwise zero. In reality, GeoDa creates a simple word file with a specific structure that contains the information about neighboring countries.

*1500 kilometer*: In principle, it is a 115*115 matrix with binary elements. If the great-circle distance between the capital cities of the two countries (row and column) is less than 1500 kilometers then the corresponding element equals one, otherwise zero. Again, the GeoDa output file is a simple word file, not a matrix.

*Inverse 1500 kilometer*: In principle, it is a 115*115 matrix. For neighboring countries (i.e. great-circle distance between their capital cities less than 1500 kilometers) the corresponding element equals 1/distance (km). Again, the GeoDa output file is a simple word file, not a matrix.

*4 nearest neighbors*: In principle, it is a 115*115 matrix with binary elements. For each country (row), its four nearest neighbors (columns) with the shortest great-circle distance between capital cities get 1, other countries (columns) get 0. Again, the GeoDa output file is a simple word file, not a matrix.

*6* and *8 nearest neighbors* are conceptually similar to *4 nearest neighbors*. Except that we assign 1 to six and eight nearest neighbors respectively. The remaining countries get zero.

It should be noted that I employed other types of spatial weight matrices such as 1, 3, and 5 nearest neighbors or 1000 kilometer distance criterion when executing the spatial BMA models. These matrices performed weaker compared to the ones used in this report. Furthermore, I did not use longer distances to build contiguity matrices because such matrices lead to situations that are not very logical. For instance, based on a 2500 kilometer distance criterion, Netherlands and Libya become neighbors. This means that their corruption levels influence each other which I suppose is not very logical. Thus, I strived to use spatial matrices that give a correct impression of the real world. Additionally, I did not use higher degrees of Knn spatial weights, such as 10, 15, or 20 nearest neighbors. The spatial econometrics literature generally prefers parsimonious spatial links. Adding too many neighbors to an analysis could wither the spatial effects and leads to insignificant spatial effects.

*Queen*, *1500 kilometer*, and *inverse 1500 kilometer* contiguity matrices have islands (i.e. neighborless countries) such as Sri Lanka and Australia. The *spatBMS* package cannot handle islands and fails to run the spatial BMA models. I manually modified these three contiguity matrices, in order to be able to execute the spatial BMA models. Table 1 provides the related information.

Table 1. Modifications made to spatial weight matrices

| Neighborless country | Queen – Added Neighbors | 1500 km - Added Neighbors | Inverse 1500 km - Added Neighbors |
|---|---|---|---|
| **Australia** | New Zealand | New Zealand | (New Zealand) 1/2325 |
| **Sri Lanka** | India | India | (India) 1/2427 |
| **Burundi** | Uganda and Kenya | - | - |
| **Djibouti** | Uganda and Kenya | Kenya | (Kenya) 1/1591 |
| **Dominican republic** | Jamaica | - | - |
| **Japan** | South Korea | - | - |
| **Jamaica** | Dominican republic and Honduras | - | - |
| **South Korea** | China and Japan | - | - |
| **Madagascar** | Mauritius and Mozambique | - | - |
| **Mauritius** | Madagascar | - | - |
| **New Zealand** | Australia | Australia | (Australia) 1/2325 |
| **Philippines** | Vietnam, Malaysia, and Indonesia | Vietnam | (Vietnam) 1/1754 |

Table 2. Descriptive statistics of spatial matrices

| | Queen | 4NN | 6NN | 8NN | 1500 km | Inverse 1500 km |
|---|---|---|---|---|---|---|
| **Minimum links** | 1 | 4 | 6 | 8 | 1 | 1 |
| **Maximum Links** | 12 | 4 | 6 | 8 | 25 | 25 |
| **Average Links** | 2.97 | 4 | 6 | 8 | 7.63 | 7.63 |
| **% non-zero links** | 2.58 | 3.48 | 5.22 | 6.96 | 6.63 | 6.63 |

### 5.4. Models

Our study consists of eight models. The first four models do not take spatial spillovers into account. They are ordinary BMA models. I used the R package *BMS* to run these models. Models 5, 6, 7, and 8 take spatial effects into account. They are the main models that our results and paper are built upon. I used the *spatBMS* package to run these models. For each of our eight models, results are obtained after 3 million iterations and discarding the first 300000 draws as burn-ins. Please see Table 3 for more information.

Table 3. Information about all eight models

| | Dependent variable | g-prior | Model |
|---|---|---|---|
| **Without spatial effects** | **CPI** | UIP | Model 1 |
| | | BRIC | Model 2 |
| | **Control of Corruption** | UIP | Model 3 |
| | | BRIC | Model 4 |
| **With spatial effects** | **CPI** | UIP | Model 5 |
| | | BRIC | Model 6 |
| | **Control of Corruption** | UIP | Model 7 |
| | | BRIC | Model 8 |

All our four spatial BMA models (i.e. models 5, 6, 7, 8) are executed using the same set of spatial weight matrices (i.e. *queen*, *4NN*, *6NN*, *8NN*, *1500 kilometer*, and *inverse 1500 kilometer*).

### 6. Results

Please see Tables 4 and 5.

### 7. Robustness checks

Throughout our research, *CPI* is the primary corruption index. *Control of Corruption* serves as an alternative index for robustness checks. Moreover, following Crespo Cuaresma and Feldkircher's (2013), I ran the BMA and spatial BMA models with two different g-priors. In particular, I used the "UIP" and "BRIC" g-priors. Please see the tutorial of the R package *BMS* for more information. Additionally, I executed the spatial BMA models with different specifications. For instance, instead of the coordinates of capital cities, I used the centroids of countries to calculate the distances and create the spatial weight matrices. I also used other Knn weight matrices such as 3, 5, and 7 nearest neighbors. Moreover, I used the raw data (without standardization) to run the spatial BMA models. In almost all of these models, the *rule of law* emerges as the sole and most important predictor of corruption after integrating out the spatial spillovers.

Figure 1. CPI scores of all the 115 sample countries in 2015

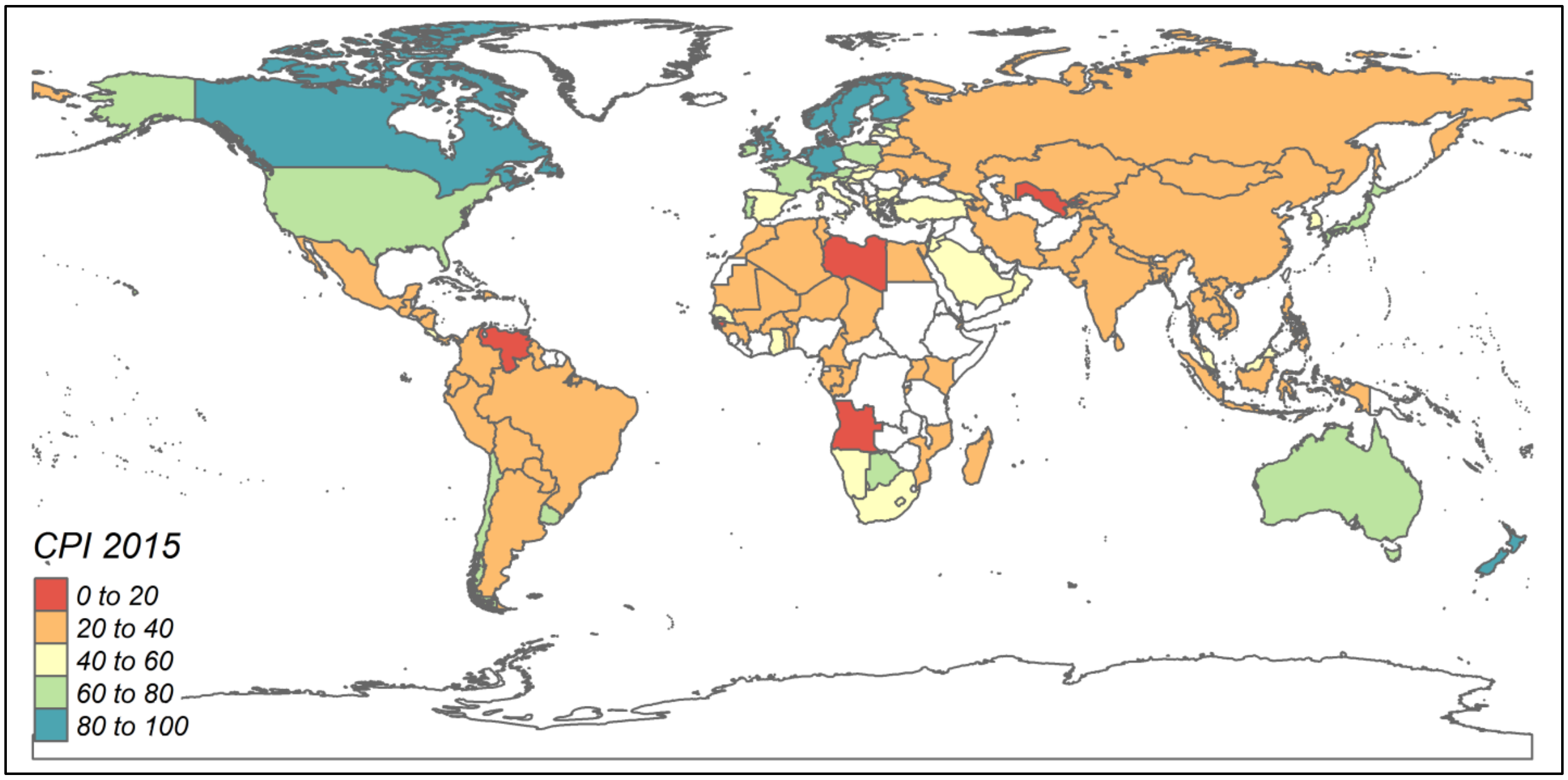

Higher CPI scores indicate lower corruption.

Table 4. BMA results, without taking spatial spillovers into account

| | CPI | | | | | | Control of Corruption | | | | | |
|---|---|---|---|---|---|---|---|---|---|---|---|---|
| | UIP | | | BRIC | | | UIP | | | BRIC | | |
| | Model 1 | | | Model 2 | | | Model 3 | | | Model 4 | | |
| | PIP | Post Mean | Post SD | PIP | Post Mean | Post SD | PIP | Post Mean | Post SD | PIP | Post Mean | Post SD |
| **Rule of law** | **1.0000** | **0.8932** | **0.0565** | **1.0000** | **0.9182** | **0.0467** | **0.9993** | **0.8254** | **0.1569** | **0.9996** | **0.8858** | **0.1252** |
| **Protestants** | **0.7885** | **0.0876** | **0.0528** | **0.6397** | **0.0713** | **0.0583** | **0.5901** | **0.0647** | **0.0596** | 0.3981 | 0.0444 | 0.0581 |
| **Urbanization** | 0.1735 | 0.0140 | 0.0334 | 0.0686 | 0.0055 | 0.0217 | 0.0181 | 0.0008 | 0.0076 | 0.0044 | 0.0002 | 0.0036 |
| **Latitude** | 0.0349 | 0.0020 | 0.0129 | 0.0093 | 0.0005 | 0.0064 | 0.0325 | 0.0022 | 0.0150 | 0.0078 | 0.0005 | 0.0069 |
| **South Asia** | 0.0302 | -0.0063 | 0.0425 | 0.0104 | -0.0023 | 0.0261 | 0.0096 | -0.0006 | 0.0155 | 0.0020 | -0.0002 | 0.0074 |
| **Colony - SP** | 0.0275 | 0.0035 | 0.0253 | 0.0068 | 0.0008 | 0.0116 | 0.0477 | 0.0080 | 0.0408 | 0.0119 | 0.0019 | 0.0195 |
| **Government effectiveness** | 0.0273 | 0.0049 | 0.0356 | 0.0083 | 0.0016 | 0.0201 | 0.2571 | 0.0874 | 0.1622 | 0.1278 | 0.0450 | 0.1254 |
| **Population** | 0.0263 | -0.0011 | 0.0086 | 0.0075 | -0.0003 | 0.0044 | 0.0183 | -0.0007 | 0.0078 | 0.0032 | -0.0001 | 0.0028 |
| **Federal** | 0.0259 | -0.0028 | 0.0212 | 0.0070 | -0.0007 | 0.0108 | 0.0185 | -0.0019 | 0.0185 | 0.0044 | -0.0004 | 0.0085 |
| **Legal - UK** | 0.0244 | -0.0024 | 0.0183 | 0.0076 | -0.0008 | 0.0103 | 0.0129 | -0.0008 | 0.0113 | 0.0033 | -0.0002 | 0.0054 |
| **Colony - UK** | 0.0217 | -0.0019 | 0.0160 | 0.0061 | -0.0006 | 0.0087 | 0.0300 | -0.0034 | 0.0227 | 0.0086 | -0.0010 | 0.0126 |
| **Legal - SOC** | 0.0211 | 0.0019 | 0.0170 | 0.0044 | 0.0004 | 0.0070 | 0.0145 | 0.0012 | 0.0145 | 0.0036 | 0.0002 | 0.0061 |
| **GDP capita** | 0.0211 | 0.0006 | 0.0114 | 0.0053 | 0.0003 | 0.0053 | 0.0112 | 0.0001 | 0.0063 | 0.0025 | 0.0001 | 0.0028 |
| **Secondary - enrolment** | 0.0176 | 0.0007 | 0.0076 | 0.0048 | 0.0002 | 0.0039 | 0.0116 | 0.0002 | 0.0054 | 0.0027 | 0.0001 | 0.0025 |
| **Latin America - Caribbean** | 0.0151 | 0.0011 | 0.0142 | 0.0034 | 0.0002 | 0.0061 | 0.0197 | 0.0020 | 0.0193 | 0.0058 | 0.0006 | 0.0101 |
| **FDI** | 0.0145 | 0.0004 | 0.0045 | 0.0036 | 0.0001 | 0.0022 | 0.0120 | 0.0002 | 0.0038 | 0.0028 | 0.0001 | 0.0019 |
| **Primary - duration** | 0.0139 | 0.0004 | 0.0061 | 0.0026 | 0.0001 | 0.0023 | 0.0239 | 0.0013 | 0.0110 | 0.0050 | 0.0003 | 0.0047 |
| **Property rights** | 0.0137 | 0.0010 | 0.0132 | 0.0030 | 0.0002 | 0.0061 | 0.0088 | 0.0000 | 0.0092 | 0.0017 | 0.0000 | 0.0038 |
| **Political rights** | 0.0137 | 0.0005 | 0.0074 | 0.0024 | 0.0001 | 0.0024 | 0.0112 | 0.0004 | 0.0072 | 0.0024 | 0.0000 | 0.0024 |
| **Colony - FR** | 0.0128 | 0.0009 | 0.0118 | 0.0027 | 0.0002 | 0.0048 | 0.0152 | 0.0012 | 0.0145 | 0.0026 | 0.0002 | 0.0057 |
| **Democracy - duration** | 0.0125 | -0.0004 | 0.0057 | 0.0027 | -0.0001 | 0.0022 | 0.0112 | -0.0003 | 0.0057 | 0.0022 | 0.0000 | 0.0020 |
| **Government size** | 0.0121 | 0.0003 | 0.0042 | 0.0028 | 0.0001 | 0.0021 | 0.0106 | 0.0001 | 0.0036 | 0.0023 | 0.0000 | 0.0016 |
| **Sub-Saharan Africa** | 0.0121 | -0.0004 | 0.0096 | 0.0027 | -0.0001 | 0.0044 | 0.0186 | -0.0016 | 0.0169 | 0.0054 | -0.0005 | 0.0089 |
| **Trade openness** | 0.0121 | -0.0003 | 0.0046 | 0.0021 | 0.0000 | 0.0015 | 0.0137 | -0.0004 | 0.0057 | 0.0027 | -0.0001 | 0.0020 |
| **Imports** | 0.0118 | -0.0002 | 0.0045 | 0.0028 | 0.0000 | 0.0017 | 0.0115 | -0.0003 | 0.0046 | 0.0022 | 0.0000 | 0.0016 |
| **Europe - Central Asia** | 0.0118 | 0.0004 | 0.0084 | 0.0022 | 0.0001 | 0.0034 | 0.0104 | 0.0001 | 0.0087 | 0.0020 | 0.0000 | 0.0034 |
| **Economic globalization** | 0.0116 | 0.0003 | 0.0060 | 0.0028 | 0.0001 | 0.0029 | 0.0102 | -0.0001 | 0.0054 | 0.0016 | 0.0000 | 0.0019 |
| **NR rents** | 0.0111 | -0.0001 | 0.0036 | 0.0022 | 0.0000 | 0.0014 | 0.0210 | -0.0009 | 0.0077 | 0.0047 | -0.0002 | 0.0034 |
| **Women in parliament** | 0.0111 | 0.0001 | 0.0040 | 0.0026 | 0.0000 | 0.0021 | 0.0628 | 0.0045 | 0.0196 | 0.0362 | 0.0029 | 0.0160 |
| **Polity 2** | 0.0110 | -0.0002 | 0.0046 | 0.0021 | 0.0000 | 0.0017 | 0.0106 | -0.0001 | 0.0049 | 0.0017 | 0.0000 | 0.0015 |
| **MENA** | 0.0109 | -0.0004 | 0.0113 | 0.0026 | -0.0001 | 0.0049 | 0.0095 | -0.0005 | 0.0111 | 0.0027 | -0.0002 | 0.0063 |
| **Press freedom** | 0.0103 | 0.0001 | 0.0057 | 0.0022 | 0.0000 | 0.0022 | 0.0122 | 0.0004 | 0.0072 | 0.0021 | 0.0000 | 0.0025 |
| **Fractionalization - language** | 0.0099 | 0.0000 | 0.0031 | 0.0018 | 0.0000 | 0.0012 | 0.0103 | -0.0001 | 0.0035 | 0.0022 | 0.0000 | 0.0016 |
| **Legal - FR** | 0.0099 | -0.0001 | 0.0068 | 0.0022 | 0.0000 | 0.0028 | 0.0111 | -0.0004 | 0.0088 | 0.0025 | -0.0001 | 0.0039 |
| **Secondary - duration** | 0.0099 | -0.0001 | 0.0031 | 0.0029 | 0.0000 | 0.0016 | 0.0185 | -0.0007 | 0.0065 | 0.0047 | -0.0002 | 0.0034 |
| **North America** | 0.0099 | 0.0008 | 0.0222 | 0.0023 | 0.0002 | 0.0104 | 0.0090 | 0.0000 | 0.0211 | 0.0019 | 0.0000 | 0.0095 |
| **Fractionalization - religion** | 0.0096 | -0.0001 | 0.0029 | 0.0025 | 0.0000 | 0.0014 | 0.0098 | -0.0001 | 0.0032 | 0.0021 | 0.0000 | 0.0014 |
| **East Asia - Pacific** | 0.0096 | -0.0001 | 0.0086 | 0.0021 | 0.0000 | 0.0039 | 0.0104 | 0.0002 | 0.0101 | 0.0020 | 0.0000 | 0.0042 |
| **Fractionalization - ethnic** | 0.0085 | 0.0000 | 0.0030 | 0.0017 | 0.0000 | 0.0013 | 0.0211 | -0.0009 | 0.0081 | 0.0045 | -0.0002 | 0.0038 |

We follow the classification used by Crespo Cuaresma and Feldkircher (2013), and Oberdabernig et al. (2018) and consider variables with PIP > 0.5 as robust determinants of corruption. In each of the four models, bold numbers indicate variables with PIP > 0.5. PIP: posterior inclusion probability. PIP indicates the importance of explanatory variables (the higher the PIP, the more important the variable in explaining corruption levels). Post mean: posterior mean. Post SD: posterior standard deviation.

Our BMA models identify the *rule of law* and *Protestants* variables as the robust determinants of corruption. The PIPs of both variables exceed 0.5 in models 1, 2, and 3. In model 4, only the *rule of law* has a PIP greater than 0.5 and could be considered as a strong regressor of corruption. The posterior coefficient (posterior mean) of *Protestants* equals 0.0647 in model 3. This means that other things being constant, a one standard deviation increase in Protestants population associates with a 0.0647 standard deviation increase in *Control of Corruption* index on average (corruption decreases).

Table 5. Spatial BMA results, taking spatial spillovers into account

| | CPI | | | | | | Control of Corruption | | | | | |
|---|---|---|---|---|---|---|---|---|---|---|---|---|
| | UIP | | | BRIC | | | UIP | | | BRIC | | |
| | Model 5 | | | Model 6 | | | Model 7 | | | Model 8 | | |
| | PIP | Post Mean | Post SD | PIP | Post Mean | Post SD | PIP | Post Mean | Post SD | PIP | Post Mean | Post SD |
| **Rule of law** | **0.9998** | **0.8005** | **0.0996** | **1.0000** | **0.8246** | **0.0705** | **0.9665** | **0.6736** | **0.2098** | **0.9804** | **0.7431** | **0.1655** |
| **Government effectiveness** | 0.1063 | 0.0309 | 0.0979 | 0.0422 | 0.0121 | 0.0621 | 0.3317 | 0.1367 | 0.2197 | 0.1612 | 0.0671 | 0.1699 |
| **Protestants** | 0.0506 | 0.0041 | 0.0199 | 0.0173 | 0.0014 | 0.0116 | 0.0317 | 0.0022 | 0.0143 | 0.0102 | 0.0008 | 0.0084 |
| **Legal - SOC** | 0.0318 | 0.0053 | 0.0338 | 0.0092 | 0.0016 | 0.0181 | 0.0150 | 0.0017 | 0.0188 | 0.0034 | 0.0004 | 0.0088 |
| **Population** | 0.0302 | -0.0016 | 0.0108 | 0.0088 | -0.0004 | 0.0053 | 0.0519 | -0.0036 | 0.0177 | 0.0115 | -0.0008 | 0.0079 |
| **Federal** | 0.0252 | -0.0032 | 0.0231 | 0.0061 | -0.0007 | 0.0110 | 0.0259 | -0.0036 | 0.0262 | 0.0057 | -0.0008 | 0.0121 |
| **Latitude** | 0.0232 | 0.0027 | 0.0211 | 0.0067 | 0.0007 | 0.0103 | 0.0192 | 0.0019 | 0.0167 | 0.0051 | 0.0005 | 0.0088 |
| **South Asia** | 0.0179 | -0.0038 | 0.0350 | 0.0053 | -0.0012 | 0.0192 | 0.0110 | -0.0014 | 0.0201 | 0.0031 | -0.0004 | 0.0109 |
| **Colony - FR** | 0.0172 | 0.0021 | 0.0197 | 0.0044 | 0.0005 | 0.0098 | 0.0281 | 0.0043 | 0.0294 | 0.0075 | 0.0011 | 0.0148 |
| **Legal - FR** | 0.0145 | -0.0016 | 0.0184 | 0.0034 | -0.0003 | 0.0077 | 0.0113 | -0.0008 | 0.0114 | 0.0029 | -0.0002 | 0.0056 |
| **Colony - UK** | 0.0144 | -0.0013 | 0.0135 | 0.0039 | -0.0003 | 0.0068 | 0.0241 | -0.0028 | 0.0216 | 0.0061 | -0.0007 | 0.0104 |
| **Legal - UK** | 0.0137 | -0.0016 | 0.0185 | 0.0031 | -0.0003 | 0.0074 | 0.0094 | -0.0006 | 0.0106 | 0.0023 | -0.0001 | 0.0049 |
| **East Asia - Pacific** | 0.0128 | 0.0023 | 0.0276 | 0.0040 | 0.0008 | 0.0155 | 0.0091 | -0.0004 | 0.0111 | 0.0018 | 0.0000 | 0.0045 |
| **MENA** | 0.0116 | -0.0017 | 0.0216 | 0.0024 | -0.0004 | 0.0096 | 0.0108 | -0.0013 | 0.0198 | 0.0030 | -0.0004 | 0.0105 |
| **Urbanization** | 0.0113 | 0.0005 | 0.0062 | 0.0034 | 0.0001 | 0.0033 | 0.0076 | 0.0000 | 0.0038 | 0.0015 | 0.0000 | 0.0016 |
| **Secondary - enrolment** | 0.0098 | 0.0005 | 0.0073 | 0.0021 | 0.0001 | 0.0034 | 0.0079 | -0.0001 | 0.0054 | 0.0018 | 0.0000 | 0.0024 |
| **North America** | 0.0093 | 0.0017 | 0.0281 | 0.0022 | 0.0004 | 0.0127 | 0.0082 | 0.0009 | 0.0248 | 0.0016 | 0.0002 | 0.0102 |
| **NR rents** | 0.0092 | -0.0002 | 0.0041 | 0.0019 | 0.0000 | 0.0017 | 0.0150 | -0.0007 | 0.0070 | 0.0042 | -0.0002 | 0.0037 |
| **Primary - duration** | 0.0089 | 0.0003 | 0.0052 | 0.0022 | 0.0001 | 0.0024 | 0.0180 | 0.0010 | 0.0095 | 0.0042 | 0.0002 | 0.0044 |
| **Government size** | 0.0088 | 0.0002 | 0.0040 | 0.0018 | 0.0001 | 0.0018 | 0.0133 | 0.0005 | 0.0057 | 0.0029 | 0.0001 | 0.0026 |
| **Europe - Central Asia** | 0.0084 | 0.0006 | 0.0134 | 0.0022 | 0.0002 | 0.0067 | 0.0111 | 0.0012 | 0.0170 | 0.0024 | 0.0003 | 0.0081 |
| **FDI** | 0.0081 | 0.0001 | 0.0028 | 0.0016 | 0.0000 | 0.0012 | 0.0080 | 0.0001 | 0.0028 | 0.0019 | 0.0000 | 0.0013 |
| **Imports** | 0.0080 | -0.0001 | 0.0034 | 0.0019 | 0.0000 | 0.0013 | 0.0093 | -0.0001 | 0.0045 | 0.0015 | 0.0000 | 0.0012 |
| **Democracy - duration** | 0.0077 | -0.0002 | 0.0048 | 0.0015 | 0.0000 | 0.0020 | 0.0081 | -0.0002 | 0.0052 | 0.0021 | -0.0001 | 0.0025 |
| **Colony - SP** | 0.0077 | 0.0005 | 0.0119 | 0.0016 | 0.0001 | 0.0049 | 0.0088 | 0.0008 | 0.0143 | 0.0021 | 0.0002 | 0.0065 |
| **Trade openness** | 0.0075 | -0.0001 | 0.0038 | 0.0015 | 0.0000 | 0.0013 | 0.0115 | -0.0004 | 0.0062 | 0.0016 | 0.0000 | 0.0017 |
| **GDP capita** | 0.0073 | 0.0002 | 0.0057 | 0.0020 | 0.0001 | 0.0029 | 0.0090 | -0.0002 | 0.0066 | 0.0021 | 0.0000 | 0.0029 |
| **Fractionalization - religion** | 0.0073 | 0.0000 | 0.0028 | 0.0017 | 0.0000 | 0.0013 | 0.0087 | 0.0002 | 0.0035 | 0.0025 | 0.0001 | 0.0019 |
| **Polity 2** | 0.0073 | 0.0000 | 0.0044 | 0.0012 | 0.0000 | 0.0016 | 0.0083 | -0.0002 | 0.0053 | 0.0024 | -0.0001 | 0.0026 |
| **Economic globalization** | 0.0071 | 0.0000 | 0.0048 | 0.0013 | 0.0000 | 0.0018 | 0.0115 | -0.0007 | 0.0101 | 0.0020 | -0.0001 | 0.0033 |
| **Women in parliament** | 0.0071 | 0.0001 | 0.0031 | 0.0020 | 0.0000 | 0.0016 | 0.0183 | 0.0009 | 0.0081 | 0.0052 | 0.0003 | 0.0045 |
| **Political rights** | 0.0070 | 0.0002 | 0.0055 | 0.0019 | 0.0001 | 0.0026 | 0.0090 | 0.0003 | 0.0064 | 0.0021 | 0.0001 | 0.0029 |
| **Property rights** | 0.0070 | 0.0002 | 0.0081 | 0.0013 | 0.0000 | 0.0032 | 0.0084 | -0.0003 | 0.0104 | 0.0019 | -0.0001 | 0.0048 |
| **Secondary - duration** | 0.0068 | -0.0001 | 0.0032 | 0.0015 | 0.0000 | 0.0014 | 0.0073 | -0.0001 | 0.0035 | 0.0018 | 0.0000 | 0.0016 |
| **Fractionalization - ethnic** | 0.0068 | 0.0000 | 0.0029 | 0.0014 | 0.0000 | 0.0012 | 0.0093 | -0.0002 | 0.0040 | 0.0016 | 0.0000 | 0.0015 |
| **Press freedom** | 0.0066 | 0.0000 | 0.0047 | 0.0014 | 0.0000 | 0.0020 | 0.0074 | 0.0001 | 0.0053 | 0.0018 | 0.0000 | 0.0024 |
| **Sub-Saharan Africa** | 0.0065 | -0.0002 | 0.0264 | 0.0015 | 0.0000 | 0.0118 | 0.0092 | 0.0008 | 0.0152 | 0.0018 | 0.0002 | 0.0064 |
| **Latin America - Caribbean** | 0.0064 | 0.0001 | 0.0143 | 0.0016 | 0.0000 | 0.0065 | 0.0077 | 0.0003 | 0.0157 | 0.0019 | 0.0000 | 0.0072 |
| **Fractionalization - language** | 0.0063 | 0.0000 | 0.0028 | 0.0013 | 0.0000 | 0.0012 | 0.0078 | 0.0001 | 0.0036 | 0.0016 | 0.0000 | 0.0016 |

We follow the classification used by Crespo Cuaresma and Feldkircher (2013), and Oberdabernig et al. (2018) and consider variables with PIP > 0.5 as robust determinants of corruption. In each of the four models, bold numbers indicate variables with PIP > 0.5. PIP: posterior inclusion probability. PIP indicates the importance of explanatory variables (the higher the PIP, the more important the variable in explaining corruption levels). Post mean: posterior mean. Post SD: posterior standard deviation.

Our spatial BMA results accentuate the significant role of the *rule of law* in keeping corruption under control. All spatial BMA models select the *rule of law* as the most important predictor of corruption. It persistently gets overwhelming posterior support which is reflected in its very high PIP scores across models 5 through 8. Based on our spatial BMA outcomes, other explanatory variables are not robust determinants of corruption. The low PIP of the remaining variables renders them as weak regressors of corruption. It is worth mentioning that the results reported in Tables 5 are obtained after controlling the spatial relationships between corruption levels across sample countries. It is interesting to pay attention to the poor performance of the *Protestants* variable in the spatial BMA context. After controlling the spatial effects, the PIP of the *Protestants* variable shrinks drastically from 0.7885 to 0.0506 (compare models 1 and 5. Other pairs of models illustrate the same pattern). This shows how neglecting the spatial relationships could result in omitted variable bias and gives way to overestimating the coefficients or overconfidence in some explanatory variables. Crespo Cuaresma and Feldkircher (2013) and Oberdabernig et al. (2018) elaborate on this issue with more details. We follow Oberdabernig et al. (2018) to interpret the posterior coefficients. For instance, based on Model 8, we can conclude that *ceteris paribus,* one standard deviation increase in rule of law associates with 0.7431 standard deviation increase in *Control of Corruption* index on average (which means decreasing corruption).

Table 6. Best spatial weight matrices. Posterior inclusion probabilities of each spatial matrix[1]

| | **Queen** | **4NN** | **6NN** | **8NN** | **1500 km** | **Inverse 1500 km** |
|---|---|---|---|---|---|---|
| **Model 5** | 0.34 | 0.02 | 0 | 99.35 | 0.15 | 0.15 |
| **Model 6** | 0.02 | 0 | 0 | 99.97 | 0 | 0 |
| **Model 7** | 0.16 | 0 | 0 | 99.79 | 0.01 | 0.01 |
| **Model 8** | 0.02 | 0 | 0 | 99.97 | 0 | 0 |

Table 6 gives the frequency with which each spatial weight matrix has been visited by the sampler. As it is evident from the table, the *8 nearest neighbor* spatial weight matrix gets the maximum posterior support across all four spatial BMA models. Thus, among the six candidate spatial matrices, the 8NN spatial weight matrix appears to be the best structure to model the spatial spillovers of corruption levels. We later use this spatial weight matrix (i.e. 8NN) to perform the Moran's I test.

1 . For each model (row) in the table, maybe the sums do not equal one due to rounding the decimals.

Figure 2. Posterior inclusion probabilities (PIPs) of explanatory variables across different spatial models

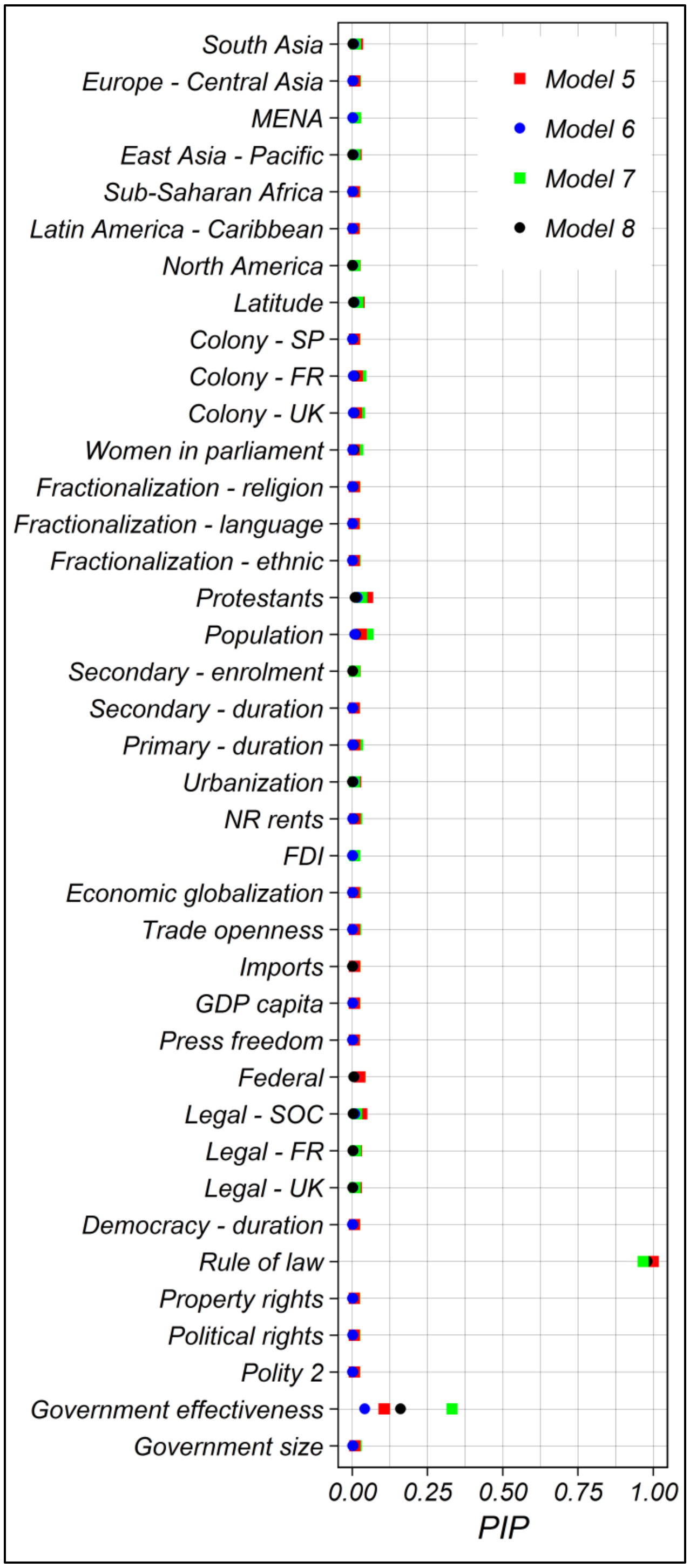

Figure 3. Posterior distributions of the five variables with the highest PIPs in the spatial BMA models, when the *g* parameter is set to UIP in spatial models

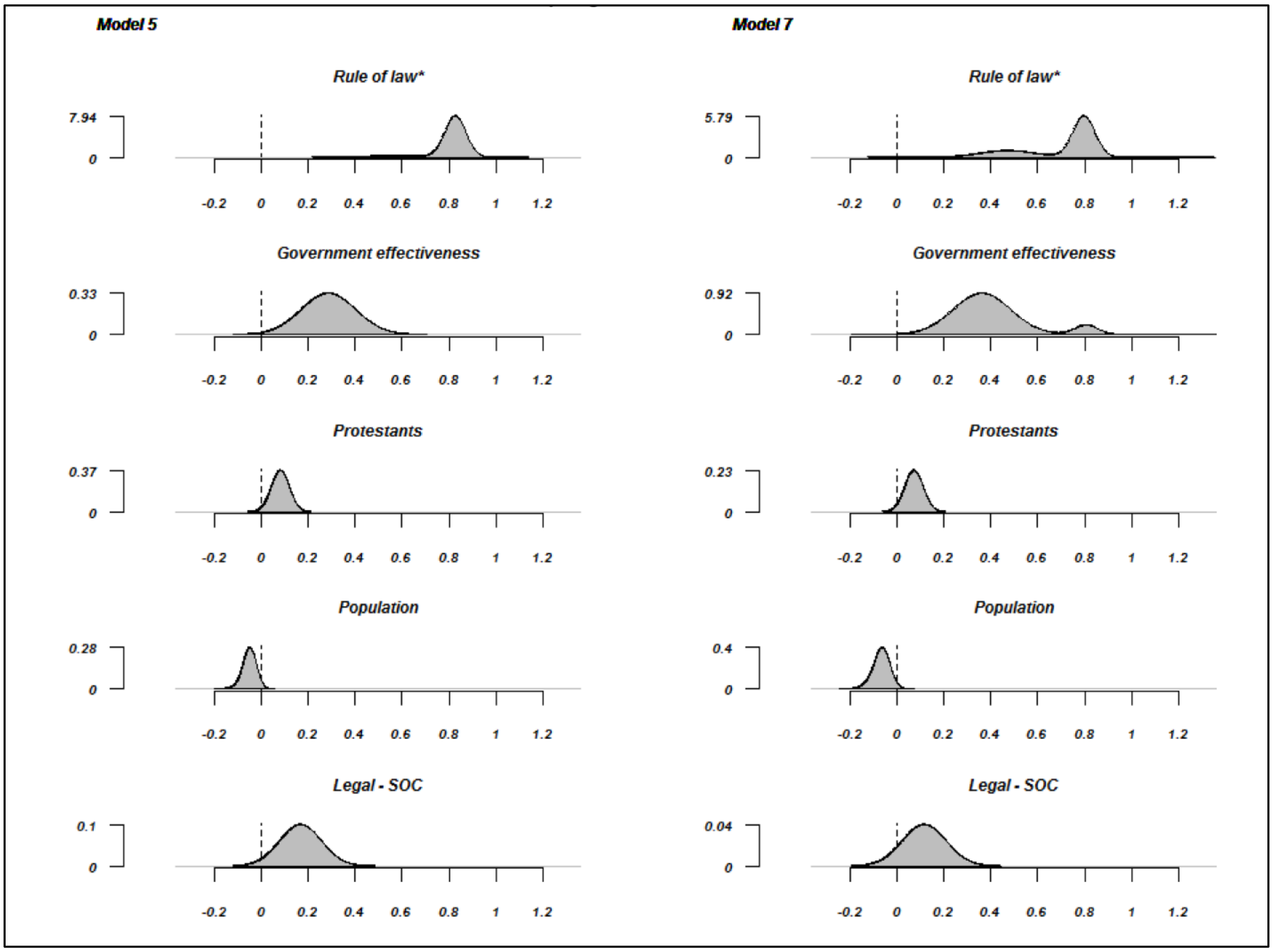

Variables with PIP > 0.5 are tagged with *. For each plot, X and Y axes present posterior coefficients and kernel density estimates respectively.

Figure 4. Posterior distributions of the five variables with the highest PIPs in the spatial BMA models, when the *g* parameter is set to BRIC in spatial models

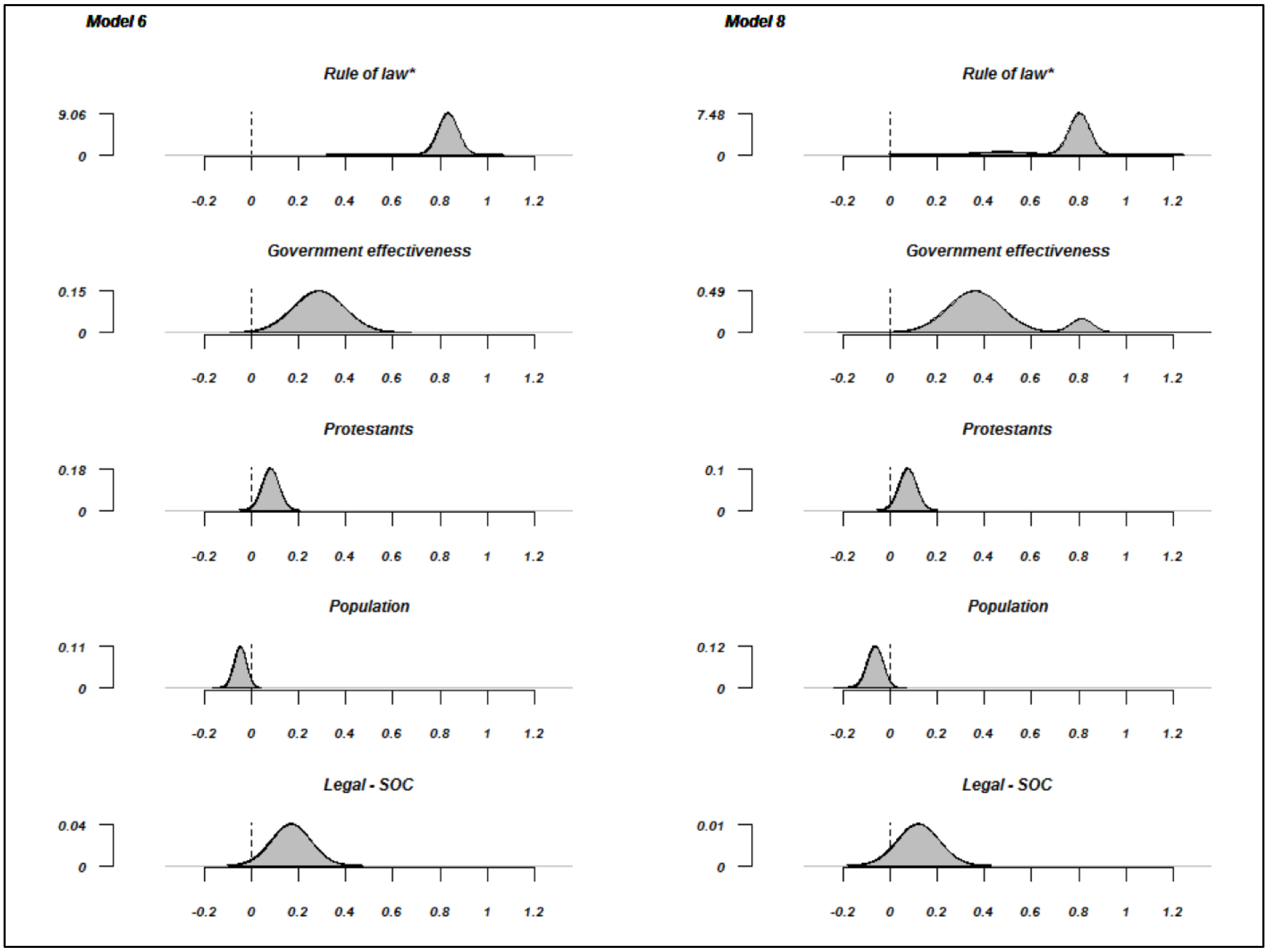

Variables with PIP > 0.5 are tagged with *. For each plot, X and Y axes present posterior coefficients and kernel density estimates respectively.

Figure 5. Moran's I test results, when the *g* parameter is set to UIP

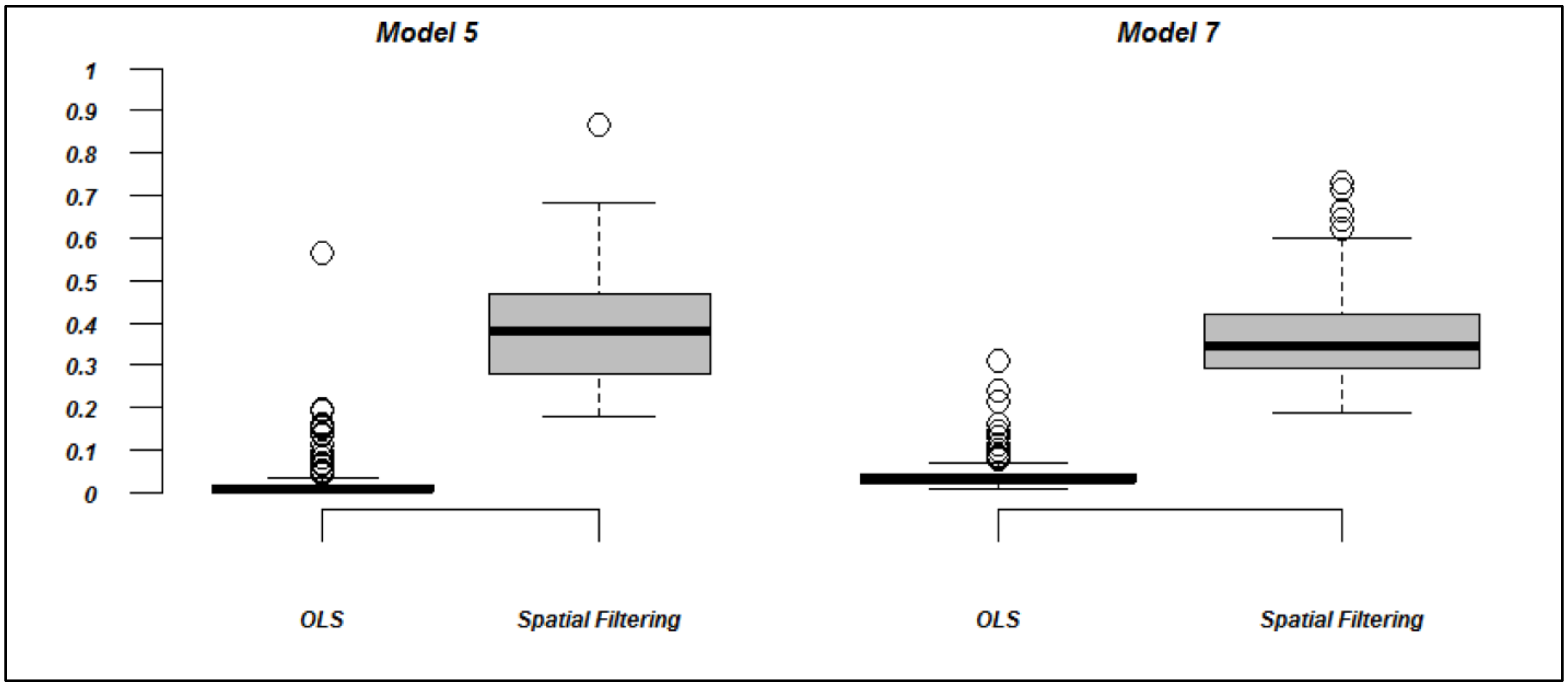

Y-axis: p-values derived from Moran's I tests.

Now, we check whether there is remaining spatial autocorrelation present in the residuals. Figure 5 compares the non-spatial regression approach and the spatial filtering BMA approach. For each model, the figure shows the distribution of the p-values of the Moran's I test, once for pure OLS regressions (without any spatial correction, left panel, OLS) and once augmented with the eigenvectors identified by the spatial filtering (right panel, Spatial Filtering). The null hypothesis corresponds to no spatial autocorrelation. As one can observe, for each model the p-values obtained from Moran's I test for OLS regressions are very close to zero which reject the null hypothesis (therefore, confirm the existence of spatial effects). After we use the mentioned spatial BMA model (model 5 or 7), the residuals show no sign of spatial patterns and as the figure depicts, the p-values of spatial filtering approach are approximately concentrated between 0.3 - 0.5 which means, we cannot reject the null hypothesis. Thus, by incorporating the eigenvectors we successfully removed spatial residual autocorrelation from the regressions in the BMA context.

Figure 6. Moran's I test results, when the *g* parameter is set to BRIC

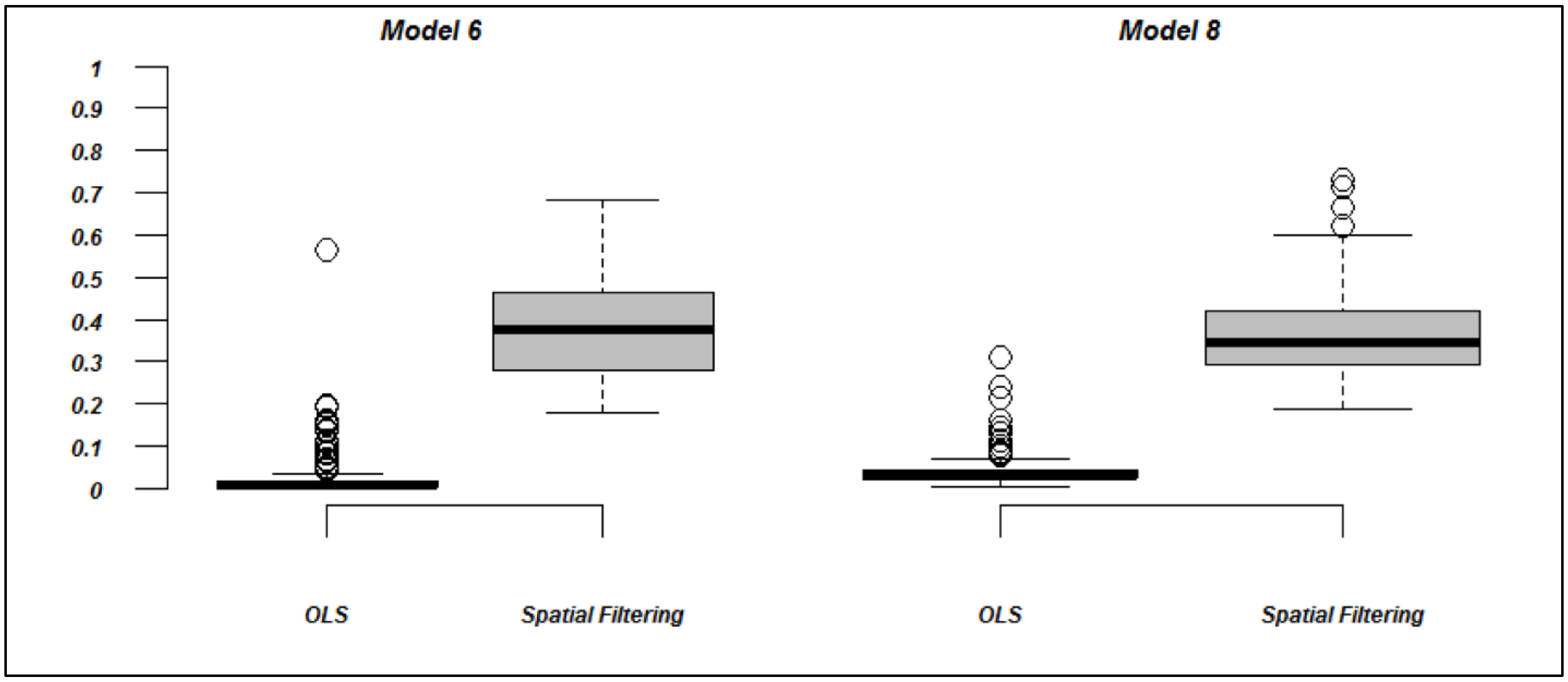

Y-axis: p-values derived from Moran's I tests.

Table 7. Summary statistics of full sample (115 countries)

| Variable | Mean | SD | Median | Min | Max | Corr. with CPI | Corr. with Control of Corruption |
|---|---|---|---|---|---|---|---|
| **CPI** | 0.00 | 1.00 | -0.37 | -1.48 | 2.36 | 1.00 | 0.99 |
| **Control of Corruption** | 0.00 | 1.00 | -0.30 | -1.56 | 2.32 | 0.99 | 1.00 |
| **Government size** | 0.00 | 1.00 | 0.09 | -2.13 | 3.85 | 0.43 | 0.41 |
| **Government effectiveness** | 0.00 | 1.00 | -0.21 | -1.48 | 2.20 | 0.95 | 0.95 |
| **Polity 2** | 0.00 | 1.00 | 0.29 | -2.25 | 1.11 | 0.54 | 0.55 |
| **Political rights** | 0.00 | 1.00 | -0.15 | -1.29 | 1.81 | -0.70 | -0.70 |
| **Property rights** | 0.00 | 1.00 | -0.24 | -1.78 | 1.99 | 0.92 | 0.91 |
| **Rule of law** | 0.00 | 1.00 | -0.35 | -1.46 | 2.05 | 0.96 | 0.95 |
| **Democracy - duration** | 0.00 | 1.00 | 0.24 | -1.35 | 1.11 | 0.57 | 0.58 |
| **Legal – UK** | 0.23 | 0.42 | 0.00 | 0.00 | 1.00 | 0.15 | 0.16 |
| **Legal – FR** | 0.48 | 0.50 | 0.00 | 0.00 | 1.00 | -0.29 | -0.30 |
| **Legal - SOC** | 0.22 | 0.41 | 0.00 | 0.00 | 1.00 | -0.15 | -0.15 |
| **Federal** | 0.15 | 0.36 | 0.00 | 0.00 | 1.00 | 0.15 | 0.16 |
| **Press freedom** | 0.00 | 1.00 | 0.08 | -1.76 | 1.86 | -0.76 | -0.75 |
| **GDP capita** | 0.00 | 1.00 | -0.05 | -1.89 | 1.90 | 0.78 | 0.76 |
| **Imports** | 0.00 | 1.00 | -0.25 | -1.68 | 3.32 | -0.03 | -0.04 |
| **Trade openness** | 0.00 | 1.00 | -0.19 | -1.67 | 2.89 | 0.07 | 0.06 |
| **Economic globalization** | 0.00 | 1.00 | -0.17 | -2.02 | 2.36 | 0.76 | 0.74 |
| **FDI** | 0.00 | 1.00 | -0.23 | -1.14 | 5.11 | 0.06 | 0.05 |
| **NR rents** | 0.00 | 1.00 | -0.44 | -0.76 | 3.45 | -0.34 | -0.38 |
| **Urbanization** | 0.00 | 1.00 | 0.13 | -2.18 | 1.96 | 0.57 | 0.55 |
| **Primary - duration** | 0.00 | 1.00 | 0.52 | -2.49 | 1.87 | 0.22 | 0.23 |
| **Secondary - duration** | 0.00 | 1.00 | -0.30 | -2.89 | 3.09 | 0.01 | -0.02 |
| **Secondary - enrolment** | 0.00 | 1.00 | 0.26 | -2.01 | 2.38 | 0.65 | 0.64 |
| **Population** | 0.00 | 1.00 | -0.11 | -1.89 | 3.11 | -0.06 | -0.04 |
| **Protestants** | 0.00 | 1.00 | -0.47 | -0.53 | 4.20 | 0.57 | 0.57 |
| **Fractionalization - ethnic** | 0.00 | 1.00 | 0.11 | -1.76 | 1.92 | -0.48 | -0.51 |
| **Fractionalization - language** | 0.00 | 1.00 | -0.13 | -1.32 | 1.90 | -0.32 | -0.33 |
| **Fractionalization - religion** | 0.00 | 1.00 | 0.09 | -1.73 | 1.91 | 0.14 | 0.14 |
| **Women in parliament** | 0.00 | 1.00 | -0.27 | -1.83 | 3.12 | 0.38 | 0.42 |
| **Colony - UK** | 0.23 | 0.42 | 0.00 | 0.00 | 1.00 | 0.16 | 0.15 |
| **Colony - FR** | 0.18 | 0.39 | 0.00 | 0.00 | 1.00 | -0.31 | -0.31 |
| **Colony - SP** | 0.14 | 0.35 | 0.00 | 0.00 | 1.00 | -0.13 | -0.12 |
| **Latitude** | 0.00 | 1.00 | -0.08 | -1.60 | 2.09 | 0.60 | 0.59 |
| **North America** | 0.02 | 0.13 | 0.00 | 0.00 | 1.00 | 0.24 | 0.23 |
| **Latin America - Caribbean** | 0.17 | 0.37 | 0.00 | 0.00 | 1.00 | -0.16 | -0.14 |
| **Sub-Saharan Africa** | 0.24 | 0.43 | 0.00 | 0.00 | 1.00 | -0.30 | -0.32 |
| **East Asia - Pacific** | 0.11 | 0.32 | 0.00 | 0.00 | 1.00 | 0.05 | 0.07 |
| **MENA** | 0.11 | 0.32 | 0.00 | 0.00 | 1.00 | -0.01 | -0.02 |
| **Europe - Central Asia** | 0.31 | 0.47 | 0.00 | 0.00 | 1.00 | 0.37 | 0.36 |
| **South Asia** | 0.04 | 0.21 | 0.00 | 0.00 | 1.00 | -0.14 | -0.11 |

Table 8. Full sample (115 countries) and corruption scores in 2015

| Country | CPI | Control of Corruption | Country | CPI | Control of Corruption | Country | CPI | Control of Corruption |
|---|---|---|---|---|---|---|---|---|
| **East Asia and Pacific (N= 13)** | | | **Chile** | 70 | 1.28 | **Mali** | 35 | -0.69 |
| **Australia** | 79 | 1.88 | **Colombia** | 37 | -0.30 | **Mauritius** | 53 | 0.30 |
| **Cambodia** | 21 | -1.12 | **Costa Rica** | 55 | 0.75 | **Mauritania** | 31 | -0.92 |
| **China** | 37 | -0.28 | **Dominican Republic** | 33 | -0.82 | **Mozambique** | 31 | -0.75 |
| **Japan** | 75 | 1.57 | **Ecuador** | 32 | -0.67 | **Niger** | 34 | -0.64 |
| **Korea, Rep.** | 54 | 0.37 | **Guatemala** | 28 | -0.73 | **Guinea-Bissau** | 17 | -1.48 |
| **Lao PDR** | 25 | -0.91 | **Guyana** | 29 | -0.64 | **South Africa** | 44 | 0.03 |
| **Mongolia** | 39 | -0.49 | **Honduras** | 31 | -0.57 | **Lesotho** | 44 | 0.07 |
| **Malaysia** | 50 | 0.24 | **Jamaica** | 41 | -0.23 | **Botswana** | 63 | 0.85 |
| **New Zealand** | 91 | 2.28 | **Mexico** | 31 | -0.77 | **Senegal** | 44 | 0.06 |
| **Philippines** | 35 | -0.45 | **Nicaragua** | 27 | -0.89 | **Togo** | 32 | -0.73 |
| **Thailand** | 38 | -0.49 | **Paraguay** | 27 | -0.95 | **Uganda** | 25 | -1.05 |
| **Vietnam** | 31 | -0.43 | **Peru** | 36 | -0.53 | **Burkina Faso** | 38 | -0.28 |
| **Indonesia** | 36 | -0.46 | **Panama** | 39 | -0.37 | **Namibia** | 53 | 0.32 |
| | | | **Uruguay** | 74 | 1.32 | | | |
| **Europe and Central Asia (N= 36)** | | | **Venezuela, RB** | 17 | -1.39 | | | |
| **Azerbaijan** | 29 | -0.93 | | | | | | |
| **Albania** | 36 | -0.48 | **MENA (N= 13)** | | | | | |
| **Armenia** | 35 | -0.53 | **Algeria** | 36 | -0.66 | | | |
| **Bulgaria** | 41 | -0.26 | **Bahrain** | 51 | 0.14 | | | |
| **Denmark** | 91 | 2.21 | **Egypt, Arab Rep.** | 36 | -0.64 | | | |
| **Ireland** | 75 | 1.62 | **Iran, Islamic Rep.** | 27 | -0.60 | | | |
| **Estonia** | 70 | 1.29 | **Israel** | 61 | 0.94 | | | |
| **Austria** | 76 | 1.52 | **Jordan** | 53 | 0.26 | | | |
| **Finland** | 90 | 2.28 | **Kuwait** | 49 | -0.23 | | | |
| **France** | 70 | 1.31 | **Libya** | 16 | -1.62 | | | |
| **Georgia** | 52 | 0.68 | **Morocco** | 36 | -0.22 | | | |
| **Germany** | 81 | 1.84 | **Oman** | 45 | 0.27 | | | |
| **Greece** | 46 | -0.08 | **Qatar** | 71 | 0.89 | | | |
| **Croatia** | 51 | 0.25 | **Saudi Arabia** | 52 | 0.05 | | | |
| **Hungary** | 51 | 0.15 | **Tunisia** | 38 | -0.07 | | | |
| **Italy** | 44 | 0.02 | | | | | | |
| **Kyrgyz Republic** | 28 | -1.15 | **North America (N= 2)** | | | | | |
| **Kazakhstan** | 28 | -0.85 | **Canada** | 83 | 1.89 | | | |
| **Latvia** | 56 | 0.47 | **United States** | 76 | 1.40 | | | |
| **Belarus** | 32 | -0.34 | | | | | | |
| **Slovak Republic** | 51 | 0.18 | **South Asia (N= 5)** | | | | | |
| **Netherlands** | 84 | 1.88 | **Bangladesh** | 25 | -0.81 | | | |
| **Norway** | 88 | 2.24 | **Sri Lanka** | 37 | -0.34 | | | |
| **Poland** | 63 | 0.67 | **India** | 38 | -0.35 | | | |
| **Portugal** | 64 | 0.96 | **Nepal** | 27 | -0.58 | | | |
| **Moldova** | 33 | -0.91 | **Pakistan** | 30 | -0.81 | | | |
| **Russian Federation** | 29 | -0.95 | | | | | | |
| **Slovenia** | 60 | 0.77 | **Sub-Saharan Africa (N= 27)** | | | | | |
| **Spain** | 58 | 0.58 | **Angola** | 15 | -1.39 | | | |
| **Sweden** | 89 | 2.24 | **Benin** | 37 | -0.56 | | | |
| **Switzerland** | 86 | 2.14 | **Congo, Rep.** | 23 | -1.20 | | | |
| **Tajikistan** | 26 | -1.13 | **Burundi** | 21 | -1.24 | | | |
| **Turkey** | 42 | -0.15 | **Cameroon** | 27 | -1.07 | | | |
| **United Kingdom** | 81 | 1.88 | **Chad** | 22 | -1.36 | | | |
| **Ukraine** | 27 | -0.98 | **Djibouti** | 34 | -0.66 | | | |
| **Uzbekistan** | 19 | -1.26 | **Gambia, The** | 28 | -0.77 | | | |
| | | | **Gabon** | 34 | -0.71 | | | |
| **Latin America and Caribbean (N= 19)** | | | **Ghana** | 47 | -0.20 | | | |
| **Argentina** | 32 | -0.55 | **Guinea** | 25 | -0.99 | | | |
| **Bolivia** | 34 | -0.70 | **Kenya** | 25 | -1.01 | | | |
| **Brazil** | 38 | -0.40 | **Madagascar** | 28 | -0.85 | | | |

Table 9. Data descriptions and sources

| Variable | Source | Definition |
|---|---|---|
| **Dependent variables** | | |
| **CPI** | Transparency International | CPI (year: 2015)<br>0: highly corrupt, 100: very clean |
| **Control of Corruption** | World Bank - WGI | Control of Corruption: Estimate (year: 2015)<br>-2.5: highly corrupt, 2.5: very clean |
| **Independent variables** | | |
| **Institutional Factors** | | |
| **Government size** | World Bank - WDI | General government final consumption expenditure (% of GDP) |
| **Government effectiveness** | World Bank - WGI | Government Effectiveness: Estimate<br>Captures perceptions of the quality of public services, the quality of the civil service and the degree of its independence from political pressures, the quality of policy formulation and implementation, and the credibility of the government's commitment to such policies.<br>-2.5: weak performance, 2.5: strong performance |
| **Polity 2** | Center for Systemic Peace | *Polity2* index<br>-10: completely autocratic, 10: total democracy |
| **Political rights** | Teorell et al. (2011) | Measures people's freedom to participate in the political process.<br>1: most free, 7: least free |
| **Property rights** | Teorell et al. (2011) | Scores the degree to which a country's laws protect private property rights and the degree to which its government enforces those laws. It ranges from 0 to 100, where 100 represents the maximum degree of protection of property rights. |
| **Rule of law** | World Bank - WGI | Rule of Law: Estimate<br>Captures perceptions of the extent to which agents have confidence in and abide by the rules of society, and in particular the quality of contract enforcement, property rights, the police, and the courts, as well as the likelihood of crime and violence.<br>-2.5: weak rule of law, 2.5: strong rule of law |
| **Democracy - duration** | Center for Systemic Peace | number of years with *Polity2* > 5 during 1985-2015<br>(based on Center for Systemic Peace classification which categorizes countries with polity score > 5 as democracies) |
| **Legal - UK** | La Porta et al. (1999) | Dummy variable: Legal origin - British |
| **Legal - FR** | La Porta et al. (1999) | Dummy variable: Legal origin - French |
| **Legal - SOC** | La Porta et al. (1999) | Dummy variable: Legal origin - Socialist |
| **Federal** | Forum of Federations | Dummy variable for federal countries |
| **Press freedom** | Freedom House | Freedom of the Press<br>0: completely free, 100: totally not free |
| **Economic Factors** | | |
| **GDP capita** | World Bank - WDI | LN(GDP per capita (constant 2010 US$)) |
| **Imports** | World Bank - WDI | Imports of goods and services (% of GDP) |
| **Trade openness** | World Bank - WDI | Exports plus Imports (% of GDP) |
| **Economic globalization** | Gygli et al. (2019) | KOF Economic Globalisation Index<br>1: minimum economic globalization<br>100: maximum economic globalization |
| **FDI** | World Bank - WDI | Foreign direct investment, net inflows (% of GDP) |
| **NR rents** | World Bank - WDI | Total natural resources rents (% of GDP) |
| **Urbanization** | World Bank - WDI | Urban population (% of total population) |
| **Primary - duration** | World Bank - WDI | Primary education, duration (years) |
| **Secondary - duration** | World Bank - WDI | Secondary education, duration (years) |
| **Secondary - enrolment** | World Bank - WDI | School enrollment, secondary (% gross) |
| **Cultural Factors** | | |
| **Population** | World Bank - WDI | LN(Population, total) |
| **Protestants** | La Porta et al. (1999) | Protestants in 1980 |
| **Fractionalization - ethnic** | Alesina et al. (2003) | Ethnic fractionalization |
| **Fractionalization - language** | Alesina et al. (2003) | Language fractionalization |
| **Fractionalization - religion** | Alesina et al. (2003) | Religious fractionalization |
| **Women in parliament** | World Bank - WDI | Proportion of seats held by women in national parliaments (%) |
| **Colony - UK** | Mayer and Zignago (2011) | Dummy variable equals one if primary colonizer was UK |
| **Colony - FR** | Mayer and Zignago (2011) | Dummy variable equals one if primary colonizer was France |
| **Colony - SP** | Mayer and Zignago (2011) | Dummy variable equals one if primary colonizer was Spain |
| **Geographical Factors** | | |
| **Latitude** | La Porta et al. (1999) | Abs(latitude of capital)/90 |
| **North America**<br>**Latin America - Caribbean**<br>**Sub-Saharan Africa**<br>**East Asia - Pacific**<br>**MENA**<br>**Europe - Central Asia**<br>**South Asia** | 7 dummies based on the World Bank regional classification | North America<br>Latin America & Caribbean<br>Sub-Saharan Africa<br>East Asia & Pacific<br>MENA<br>Europe & Central Asia<br>South Asia |